\documentstyle[twoside,fleqn,espcrc2]{article}

\newcommand{\AmS}{{\protect\the\textfont2
  A\kern-.1667em\lower.5ex\hbox{M}\kern-.125emS}}

\title{KPZ Formulas for Weyl-Invariant Induced Gravity
and Topologically Massive Gravity
\thanks{Talk given at the 2nd Topical Meeting on Constrained Dynamics and 
Quantum Gravity, Santa Margherita, Ligure, Italy, 17-21 September, 1996.
To be published in the proceedings (Editor: V. de Alfaro). $~~~~~~~~~~~~~$ 
$~~~~~~~~~~~~~$ \space\space\space\space\space\space
OUTP-96-72P $~~~~~~~~~~~~~~~~~~~~~~~~~~~~~~~~~~~~~~~~~~~~~~~~~~~$
\space\space\space\space\space\space
hep-th/9701161 $~~~~~~~~~~~~~~~~~~~~~~~~~~~~~~~~~~~~~~~~~~~~~~~~~~~$
\space\space\space\space\space\space
December 1996}
}

\author{Giovanni Amelino-Camelia\address{Theoretical Physics, University of
Oxford\\
1 Keble Road, Oxford OX1 3NP, U.K.}}
       
\begin{document}

\begin{abstract}
I discuss the applicability in 
Weyl-invariant induced gravity
and
topologically massive gravity
of certain formulas
originally derived by
Knizhnik, Polyakov, and Zamolodchikov 
in the context of 
diffeomorphism-invariant induced gravity.
\end{abstract}

\maketitle

Over the last ten years there has been 
a strong interest in the investigation 
of low-dimensional quantum gravity toy models.
A very popular such model is 
the two-dimensional (2D) ``induced'' gravity
characterized by 
a diffeomorphism-invariant
measure \cite{kpz} 
for the functional (path) integration over the
metric tensor field 
and
the action 
\begin{eqnarray}
S(g) \! = {d \over 96 \pi} \!
\int \! d^2\xi_1 \, d^2\xi_2 \, \sqrt{g(\xi_1)} \, 
R(g(\xi_1)) ~~~~~~~
\nonumber\\
~~~~ \cdot \Box^{-1}(\xi_1,\xi_2) ~ \sqrt{g(\xi_2)} \, R(g(\xi_2))
\, , \label{sgp}
\end{eqnarray}
where 
$g$ is the metric tensor, $\sqrt{g}$ is the square root of the 
determinant of $g$,
$\Box^{-1}$ is the inverse of the Laplace-Beltrami operator,
and,
for simplicity, I set the {\it cosmological constant} to zero.
This can be seen as the quantum gravity induced 
by integrating out,
with a diffeomorphism-invariant
path-integral measure \cite{kpz},
the ``matter" degrees of freedom
of the field theory with action 
\begin{eqnarray}
{\cal I}(X,g) = {1 \over 2} \int d^2\xi ~ \sqrt{g} ~
g^{\mu \nu} ~ \partial_\mu X^A \, \partial_\nu X^A ~,
\label{sxg}
\end{eqnarray}
where the $X^A$, with $A \! = \! 1,2,...,d$,
denote the $d$ components of the massless ``matter" scalar field.
Note that the action (\ref{sxg}) is invariant under both diffeomorphisms
and Weyl transformations, while (\ref{sgp}) is 
only diffeomorphism-invariant.
This reflects the presence of an anomaly, {\it i.e.} the fact 
that there are no path-integral measures with invariance under
both diffeomorphisms
and Weyl transformations \cite{kpz,abs1}.

A well-appreciated result, first derived by
Knizhnik, Polyakov, and Zamolodchikov \cite{kpz},
concerns the Green's functions of 
diffeomorphism-invariant induced gravity.
It is encoded in the following anomalous Ward identities
of the light-cone-gauge formulation of the
theory (in which the entire physical content of the metric tensor
resides in its $g_{++}$ component):
\begin{eqnarray}
\sum^n_i  \! \left < g_{++} \!  (\xi_1 ) \! 
\dots \! \delta_{\! \delta \! f} \! g_{++} \! (\xi_i ) \! 
\dots  \! g_{++} \! (\xi_n)  \right > \nonumber\\
~~~ = \! 
\frac{28 \! - \! d  \! -  \! \lambda}{i 48 \pi}
 \! \! \int \!  \! d\xi^2  \, \delta \! f \! ( \! \xi \! )
~~~~~~~~~~~~~~~~~~
\nonumber\\
~~~~ \! 
\cdot
\left<  \partial_-^3 g_{++} \! (  \xi  )
g_{++} \! (\xi_1 ) \! \dots  \! g_{++} \! 
(\xi_n)  \right > 
\, , \label{Ward}
\end{eqnarray} 
where $\delta_{\! \delta \! f}$ is such that
\begin{eqnarray}
\delta_{\! \delta \! f} \chi 
= (2 \partial_+ - \chi \partial_-) \, \delta \! f +
\delta \! f \, \partial_- \chi 
\label{variation}
\end{eqnarray} 
and 
$\lambda$ is a constant characterizing the measure \cite{kpz,abs1}
that can be expressed in terms of the number of 
components of the massless ``matter" scalar field
using the relation \cite{kpz}
\begin{equation}
\label{lambda}
\lambda = {43 - d - \sqrt{(d -1)(d - 25)} \over 2}~.
\end{equation}

Another important observation due to
Knizhnik, Polyakov, and Zamolodchikov \cite{kpz}
is frequently exploited in the
conformal-gauge formulation of the
theory, which is Liouville theory.
It establishes the
scaling relations \cite{kpz} for primary fields
of Liouville theory 
\begin{equation}
{\hat \Delta}-{\hat \Delta}_0=\frac{{\hat \Delta}(1-{\hat \Delta})}{c+2}
\label{KPZtransfLiouville}\end{equation}
where $c$ is the central charge of the
$SL(2, R)$ current algebra,
${\hat \Delta_0}$ is the
bare conformal dimension of the primary field,
and ${\hat \Delta}$ is its Liouville-dressed conformal dimension.

Although these results were originally derived in
diffeomorphism-invariant induced gravity,
it has been recently realized that they have wider applicability.
Let us start by considering
Weyl-invariant induced gravity \cite{kar94,jac95,abs1,reuter}, 
which is the 2D quantum gravity induced 
by integrating out,
with a Weyl-invariant
path-integral measure \cite{jac95,abs1}, 
the ``matter" degrees of freedom
of the field theory with action (\ref{sxg}).
The action governing the
dynamics of the metric tensor field is 
\begin{eqnarray}
{\cal S}(\gamma) \! = \! {d \over 96 \pi}
\int d^2\xi_1 \, d^2\xi_2 ~  \, R(\gamma(\xi_1)) ~~~~~~~~
\nonumber\\
~~~\cdot 
\Box^{-1}(\xi_1,\xi_2) \,  R(\gamma(\xi_2)) ~,
\label{sgj}
\end{eqnarray}
where I used the short-hand
notation $\gamma \! \equiv \! \sqrt{g} \, g$.
A Weyl-invariant 
measure for the functional (path) integration over the
metric tensor field $g$, depending on $g$ only through $\gamma$,
has also been identified \cite{kar94,jac95,abs1}.
The fact that both the measure and (\ref{sgj})
depend on $g$ only through $\gamma$
encodes the fact that the theory is Weyl invariant and invariant
under diffeomorphisms of unit Jacobian
(also called ``S-diffeomorphisms" 
or ``area-preserving diffeomorphisms" \cite{kar94,jac95,abs1}),
while the theory is not fully diffeomorphism-invariant.
Concerning the Green's functions of this
Weyl-invariant induced gravity
it has been recently realized \cite{abs1} that
in an appropriate (light-cone-like) gauge
the following relations hold
\begin{eqnarray}
\sum^n_i  \! \left < \gamma_{++} \!  (\xi_1 ) \! 
\dots \! \delta_{\! \delta \! f} \! \gamma_{++} \! (\xi_i ) \! 
\dots  \! \gamma_{++} \! (\xi_n)  \right > 
\nonumber\\
~~~ = \! 
\frac{28  \! \! - \! \!  d  \! \! 
-  \! \! \lambda^W}{i 48 \pi}
 \! \! \int \!  \! d\xi^2  \! \, \delta \! f \! ( \! \xi \! )
~~~~~~~~~~~~~~~~~~
\nonumber\\
~~~~~~ \! 
\cdot
\left<  \partial_-^3 \gamma_{++} \! (  \xi  )
\gamma_{++} \! (\xi_1 ) \! \dots  \! \gamma_{++} \! 
(\xi_n)  \right > 
~, \label{Wardj}
\end{eqnarray} 
where $\lambda^W$ is a constant characterizing the
Weyl-invariant measure\footnote{This constant $\lambda^W$
has been conjectured \cite{abs1} to be equal to the
constant $\lambda$ of the diffeomorphism-invariant case.}
and 
I denoted with $\gamma_{++}$ the component of $\gamma$
that is relevant in the given gauge \cite{abs1}.
The correspondence between (\ref{Ward}) and (\ref{Wardj})
describes the applicability of the KPZ formula (\ref{Ward})
in the context of Weyl-invariant induced gravity.

Actually, it has been realized in Ref.~\cite{as1} that 
a similar argument applies also to other 2D induced-gravity
theories.
In particular, a one-parameter family of such theories, all invariant 
under diffeomorphisms of unit Jacobian but not fully
diffeomorphism-invariant, has been identified, and evidence has been
found \cite{as1} in support of the applicability of 
the KPZ formula (\ref{Ward}) in those theories.
I concentrate here on the special Weyl-invariant case
for simplicity.

Let us now move on to a 3D context and consider a theory
in which topologically massive gravity
is coupled to topologically massive electrodynamics
and a charged massive scalar field.
It has been recently realized \cite{a-cks}
that the KPZ formula (\ref{KPZtransfLiouville})
encodes (in the sense that I clarify in the following)
the relation between
the Aharonov-Bohm amplitude of such a theory
and the ordinary Aharonov-Bohm amplitude \cite{ab}.
This is encoded formally in a property of the
transmuted spin $\Delta$, 
which is defined as the overall coefficient of the 
Aharonov-Bohm amplitude \cite{a-cks}
\begin{equation}
\Delta \frac{16\pi}{i} \dim({\cal G})
\frac{\epsilon_{\mu\nu\lambda}p_1^\mu p_2^\nu q^\lambda}{q^2}
~,
\label{abrelgen}\end{equation}
where $p_1$ and $p_2$ are the
initial momenta 
of the two charged particles
involved in the scattering,
$q$ is the momentum transfer,
and ${\cal G}$ is the gauge group.
For example, in the ordinary (abelian ${\cal G}$ and
no gravitational degrees of freedom)
Aharonov-Bohm amplitude
the transmuted spin is given by the inverse of 
the Chern-Simons coefficient.

In Ref.~\cite{a-cks} evidence is provided in support
of the relation 
\begin{equation}
\Delta-\Delta_0=\frac{\Delta(\Delta - 1)}{k'+2}
\label{KPZtransf}
\end{equation}
between
the {\it bare} transmuted spin $\Delta_0$ ({\it i.e.} the value 
taken by the transmuted spin 
when the ordinary gauge degrees of freedom are present, but the 
gravitational degrees of freedom are turned off),
the {\it gravitationally-dressed} transmuted 
spin $\Delta$ ({\it i.e.} the value taken by the transmuted spin 
when also the gravitational degrees of freedom 
of topologically massive gravity are turned on),
and the gravitational Chern-Simons
coefficient $k'$.

The formula (\ref{KPZtransf}) reproduces
the KPZ formula (\ref{KPZtransfLiouville})
upon identifying
the bare (respectively the Liouville-dressed)
conformal dimensions of primary fields in Liouville theory
with the bare (respectively the 
gravitationally-dressed)
transmuted spin of topologically massive gravity,
and assuming the relation $c=-k'-4$
between the central charge $c$ of Liouville theory
and the gravitational Chern-Simons
coefficient $k'$.

This observation provides one of the strongest pieces
of evidence in support of a relation of the type discussed
in Refs.~\cite{carlip3,gravrel}
between topologically massive gravity and
Liouville theory,
and can be used to motivate a topological membrane approach to string
theory \cite{kogan1}, in which the string world-sheet
is filled in and viewed as the boundary of a three-manifold.

Similarly, the applicability of KPZ formulas in
non-diffeomorphism-invariant induced-gravity theories
can be used as a probe of the relation among
the different 
2D induced gravity theories (diffeomorphism-invariant or not). 
In particular, it has been suggested \cite{abs1,as1}
that the members of a large class 
of such theories might all be essentially
equivalent (as originally established for diffeomorphism-invariant
induced gravity) to Liouville theory.
Interestingly, this has led to a different 
viewpoint \cite{bhs,as1,ags1,mariano}
on the phenomenon of Hawking radiation in 2D gravity.
The presence of a (non-necessarily-Weyl) anomaly 
and the residual invariance under
diffeomorphisms of unit Jacobian are now seen
as the essential elements of this phenomenon,
whereas in earlier interpretations the presence of a Weyl anomaly
was often emphasized.

The recent results on 2D induced gravity theories 
have also sparked renewed interest
in the issues associated to 
the quantization of anomalous field theories.
In particular, effort has been devoted  \cite{ags1,msy} 
to the investigation of the physical implications of 
local counter-terms in anomalous field theories.
While the mass emergent 
in the chiral Schwinger model \cite{jac85} 
is known to depend on the coefficient of
one such local counter-term 
(associated to a one-parameter family 
of chiral symmetry breaking measures),
similar phenomena have not been identified in the context
of bosonic induced-gravity theories, at
least in the case of topologically-trivial two-manifolds
(the only case seriously investigated in the related studies).
However, some evidence of
physical implications of 
local counter-terms 
has been found in the context of 
chiral 2D induced gravity \cite{ags1}.

\section*{ACKNOWLEDGEMENTS}
This work is based on results obtained in Refs.~\cite{abs1,as1,ags1}
in collaboration with D.~Bak, L.~Griguolo, and D.~Seminara
and in Ref.~\cite{a-cks}
in collaboration with I.I.~Kogan and R.J.~Szabo.
I would like to thank V.~de Alfaro
and all those involved in the organization of the very successful
2nd Topical Meeting 
on Constrained Dynamics and Quantum
Gravity.
I would also like to acknowledge very stimulating discussions 
with several participants to the Meeting, particularly 
M.~Cadoni, S.~Carlip, S.~Deser, R.~Jackiw, S.~Sciuto, 
and G.~Veneziano.
This work was supported in part by PPARC
and by the Foundation Stiftelsen Blanceflor Boncompagni-Ludovisi.


\begin{thebibliography}{000}

\bibitem{kpz} A.M. Polyakov, Phys. Lett. B103 (1981) 207;
Mod. Phys. Lett. A2 (1987) 893; 
V.G. Knizhnik, A.M. Polyakov and A. B. Zamolodchikov, 
{\it ibid.} A3 (1988) 819; 
F. David, {\it ibid.} A3 (1988) 1651; 
J. Distler and H. Kawai, Nucl. Phys. B321 (1989) 509;
N.E. Mavromatos and J.L. Miramontes,
Mod. Phys. Lett. A4 (1989) 1847;
E. D'Hoker and P.S. Kurzepa, 
{\it ibid.} A5 (1990) 1411.

\bibitem{abs1} G. Amelino-Camelia, D. Bak, and D. Seminara,
Phys. Lett. B354 (1995) 213.

\bibitem{kar94} D.R. Karakhanian, R.P. Manvelian, R.L. Mkrtchian,
Phys. Lett. B329 (1994) 185.

\bibitem{jac95} R. Jackiw, 
Report No. hep-th/9501016.

\bibitem{reuter} Weyl invariance is discussed within
a different formalism in M. Reuter, Report No. hep-th/9612158,
and references therein.

\bibitem{as1} G. Amelino-Camelia and D. Seminara,
Class. Quant. Grav. 13 (1996) 881.

\bibitem{a-cks} G. Amelino-Camelia, I.I. Kogan, and R.J. Szabo, 
Nucl. Phys. B480 (1996) 413.

\bibitem{ab} Y. Aharonov and D. Bohm, Phys. Rev. 115 (1959) 485.

\bibitem{carlip3} S. Carlip, Nucl. Phys. B362 (1991) 111.

\bibitem{gravrel} S. Carlip and I.I. Kogan, Phys. Rev. Lett. 64 (1990) 148;
{\it ibid.} 67 (1991) 3647; Mod. Phys. Lett. A6 (1991) 171;
S. Carlip, Phys. Rev. D45 (1992) 3584;
I.I.  Kogan, Phys. Lett. B256 (1991) 369;
I.I. Kogan, Nucl. Phys. B375 (1992) 362; 
M.C. Ashworth, Report No. hep-th/9510192.

\bibitem{kogan1} I.I. Kogan, Phys. Lett. B231 (1989) 377.

\bibitem{bhs} J. Navarro-Salas, M. Navarro, and C.F. Talavera,
Phys. Lett. B356 (1995) 217;
J. Cruz, J. Navarro-Salas, M. Navarro, and C.F. Talavera, 
Report No. hep-th/9606097.

\bibitem{ags1} G. Amelino-Camelia, L. Griguolo, and D. Seminara, 
Phys. Lett. B371 (1996) 41.

\bibitem{mariano} M. Cadoni, Report No. hep-th/9610201.

\bibitem{msy} M. Martellini, M. Spreafico, and K. Yoshida,
Phys. Rev. D54 (1996) 7664.

\bibitem{jac85} R. Jackiw and R. Rajaraman, 
Phys. Rev. Lett. 54 (1985) 1219;
for a review see R. Jackiw in {\it Quantum Mechanics of 
Fundamental Systems 1}, C. Teitelboim ed. (Plenum, New York, 1988).

\end{thebibliography}
\end{document}